\documentclass[
amsmath,
prd,nofootinbib,floatfix,12pt,
]{revtex4}

\usepackage{hyperref}

\newcommand\beq{\begin{eqnarray}}
\newcommand\eeq{\end{eqnarray}}

\def\lsim{\mathrel{\rlap{\lower4pt\hbox{$\sim$}}
    \raise1pt\hbox{$<$}}}                
\def\gsim{\mathrel{\rlap{\lower4pt\hbox{$\sim$}}
    \raise1pt\hbox{$>$}}}            

\def\MSbar{$\overline{\rm MS}$ }
\allowdisplaybreaks
\usepackage{graphicx}
\usepackage{setspace}
\usepackage{dcolumn}
\usepackage{bm}
\usepackage{verbatim}

\usepackage[dvips]{color}
\definecolor{Red}{cmyk}{0,1,1,0}
\definecolor{BrickRed}{cmyk}{0,0.89,0.94,0.28}
\definecolor{Blue}{cmyk}{1,1,0,0}
\definecolor{Green}{cmyk}{1,0,1,0}

\begin{document}

\title{\large \baselineskip=20pt 
The Standard Model at 200 GeV}

\author{Zamiul Alam and Stephen P.~Martin}
\affiliation{\it 
Department of Physics, Northern Illinois University, DeKalb IL 60115}

\begin{abstract}\normalsize \baselineskip=15.5pt
The Standard Model can be defined quantitatively by running parameters in a mass-independent renormalization scheme at a fixed reference scale. We provide a set of simple interpolation formulas that give the fundamental Lagrangian parameters in the \MSbar scheme at a renormalization scale of 200 GeV, safely above the top-quark mass and suitable for matching
to candidate new physics models at very high mass scales using renormalization group equations. These interpolation formulas take as inputs the on-shell experimental quantities, and use the best available calculations in the pure \MSbar scheme. They also serve as an accounting of the parametric uncertainties for the short-distance Standard Model Lagrangian. We also include an interpolating formula for the $W$ boson mass.
\end{abstract}

\maketitle

\tableofcontents
\baselineskip=16.7pt
\newpage

\section{Introduction \label{sec:intro}}
\setcounter{equation}{0}
\setcounter{figure}{0}
\setcounter{table}{0}
\setcounter{footnote}{1}

The Standard Model of fundamental particle physics has reached a high level of experimental maturity with the 2012 discovery of the Higgs boson and the increasingly accurate measurements of its mass, production modes, and decays. Meanwhile, the explorations of the Large Hadron Collider (LHC) at the high-energy frontier have not revealed any substantial and lasting deviations that would compel extensions of the Standard Model, despite many motivations for such new physics. 

Given this state of affairs, it is useful to summarize as accurately as possible our quantitative knowledge of the Standard Model in terms of the Lagrangian parameters that define the theory. These defining parameters can then be matched to a larger set of parameters in candidate new physics theories characterized by large mass scales. It is convenient to use the \MSbar
renormalization scheme 
\cite{Bardeen:1978yd,Braaten:1981dv}
based on dimensional regularization 
\cite{Bollini:1972ui,Ashmore:1972uj,Cicuta:1972jf,tHooft:1972fi,tHooft:1973mm}
for this purpose. The Standard Model \MSbar Lagrangian parameters to be evaluated include:
\beq
\mbox{Higgs sector:}&\>& \lambda, m^2,
\nonumber
\\
\mbox{gauge couplings:}&\>& g_3,\, g,\, g',
\nonumber
\\
\mbox{quark Yukawa couplings:}&\>& y_t,\, y_b,\, y_c,\, y_s,\, y_u,\, y_d, 
\nonumber
\\
\mbox{lepton Yukawa couplings:}&\>& y_\tau,\, y_\mu,\, y_e.
\label{eq:MSbarparameters}
\eeq
where we have neglected
the neutrino sector. Also omitted here are the four physical parameters (three flavor-mixing angles and one CP-violating phase angles) associated with the Cabibbo-Kobayashi-Maskawa (CKM) matrix for quarks, which can be considered separately and decouple from the discussion below to a high degree of accuracy due to unitarity of the CKM matrix. For a discussion, and numerical values of these four CKM angle parameters in the Standard Model, see the relevant section of the Review of Particle Properties (RPP) \cite{ParticleDataGroup:2022pth}  published by the Particle Data Group (PDG),
and references therein.

Each of the 14 quantities in eq.~(\ref{eq:MSbarparameters}) is a running parameter, dependent on the choice of \MSbar renormalization scale $Q$, governed by renormalization group equations that are now known \cite{MVI}-\cite{Davies:2019onf} with some effects through 5-loop order. The normalization convention of the Higgs sector parameters is the now-standard one such that the tree-level potential for the canonically normalized real neutral component of the Higgs field is $V = \frac{1}{2} m^2 H^2 + \frac{\lambda}{4} H^4$. In particular, the Higgs squared mass parameter $m^2$ is negative, as required by electroweak symmetry breaking, and can be traded for the vacuum expectation value (VEV) $v$ for $H$, defined as the minimum of the all-orders effective potential in Landau gauge, so that the sum of all tadpole diagrams (including the tree-level tadpole) simply vanishes. In practice, the effective potential is known fully at 2-loop \cite{Ford:1992pn,Martin:2001vx} and 3-loop \cite{Martin:2013gka,Martin:2016bgz,Martin:2017lqn} orders, supplemented by the QCD 4-loop contributions \cite{Martin:2015eia}. 
 
For the purposes of matching to ultraviolet new physics proposals, one should work in a non-decoupling scheme, in which all of the Standard Model particles including the top quark
are  propagating degrees of freedom. For this reason, we choose
as a benchmark \MSbar renormalization scale the value $Q = 200$ GeV, which is somewhat arbitrary but has the advantages of being a round number, safely above the top-quark mass, and probably\footnote{This is certainly true for theories such as supersymmetry, where current bounds on squarks and gluinos are now generally well above 1 TeV. However, in other contexts it is not guaranteed, despite the LHC's negative search results, since the new physics might be only very weakly coupled to the Standard Model.} well below the scale of new physics. Furthermore, taking a fixed scale (rather than, say, the experimental top-quark mass, which is subject to uncertainty and change) provides for better numerical stability.
The results given here can then be evolved to any desired matching scale by the Standard Model renormalization group equations. 

The standard reference for important experimental results in high energy physics, the RPP \cite{ParticleDataGroup:2022pth} published by the PDG, instead (so far, at least) summarizes our knowledge of the Standard Model
in terms of what we will refer to as the on-shell quantities. The RPP quantities that are in the most direct correspondence to the Lagrangian parameters in eq.~(\ref{eq:MSbarparameters}) are:
\beq
\mbox{fine-structure constant:}&\>& \alpha = 1/137.035999084\ldots\>\mbox{and}\>\, \Delta\alpha^{(5)}_{\rm had}(M_Z),
\nonumber
\\
\mbox{Fermi decay constant:}&\>&G_F 
,
\nonumber
\\
\mbox{5-quark QCD coupling:}&\>& \alpha_S^{(5)}(M_Z),
\nonumber
\\
\mbox{heavy particle physical masses:}&\>& M_t,\, M_h,\, M_Z,\, M_W,
\nonumber
\\
\mbox{running light quark masses:}&\>& m_b(m_b),\, m_c(m_c),\, m_s(\mbox{2 GeV}),\, m_u(\mbox{2 GeV}), \, m_d(\mbox{2 GeV}),
\phantom{xxx}   
\nonumber
\\
\mbox{lepton pole masses:}&\>& M_\tau,\, M_\mu,\, M_e.
\label{eq:onshellparameters}
\eeq
It should be noted that in this paper $M_Z$ and $M_W$ are the on-shell masses in the PDG parameterization; these are related to the gauge-invariant complex pole squared masses 
$s_p = (M_p - i \Gamma_p/2)^2$ by 
$M = M_p (1 + \delta)/\sqrt{1 - \delta}$, where $\delta = \Gamma_p^2/4 M_p^2$ in each case.
(For recent discussions, see refs.~\cite{Willenbrock:2022smq} and \cite{Martin:2022qiv}.)
In principle, the hadronic contribution to the fine-structure constant, $\Delta\alpha^{(5)}_{\rm had}(M_Z)$ is not independent, and could be determined in terms of the other quantities, but in practice it cannot be perturbatively evaluated and therefore is taken as an independent experimental input. The $W$ boson mass, $M_W$, can also be determined in terms of the others. The other 14 on-shell quantities in eq.~(\ref{eq:onshellparameters}) are dual to those of the 14 independent \MSbar parameters in eq.~(\ref{eq:MSbarparameters}). This means that one can take the \MSbar parameters as theoretical inputs with the on-shell quantities as outputs, or one can take the on-shell quantities as experimental inputs and view the \MSbar parameters as the outputs.

A great deal of effort has gone into relating the two sets of parameters, and to evaluating $M_W$ in terms of the others. For a necessarily incomplete set of earlier references, see 
\cite{Weinberg:1980wa}-\cite{Huang:2020hdv} and papers discussed therein. The present work is based on the computer code {\tt SMDR} \cite{Martin:2019lqd,SMDRWWW}, which implements calculations in refs.~\cite{Martin:2014cxa,Martin:2015lxa,Martin:2015rea,Martin:2016xsp,Martin:2018yow,Martin:2022qiv} in the tadpole-free pure \MSbar scheme. This code contains command-line utilities and library programs for fitting the on-shell parameters in terms of the \MSbar parameters, and vice versa, and for performing the renormalization group running in the Standard Model, implementing the state-of-the-art calculations. 

The purpose of the present paper is to provide simple and convenient interpolation formulas that accurately provide the \MSbar parameters in eq.~(\ref{eq:MSbarparameters}) and $M_W$ in terms of the on-shell parameters in eq.~(\ref{eq:onshellparameters}), obtained by doing a fit to the results\footnote{More specifically, the interpolation formulas in the present paper have a similar functionality to the {\tt SMDR} command-line invocation {\tt calc\_fit -Q 200}, but with a drastically shorter evaluation time.}
of {\tt SMDR}.
The simplicity and accuracy of the interpolation formulas is aided by the fact that the allowed parameters in the Standard Model are now all experimentally restricted to rather narrow ranges. 

We now discuss the organization and notation of the interpolation formulas below. First,
define a set of benchmark (denoted by subscript 0) on-shell inputs, using the values from the most recent 2022 PDG data:
\beq
&&
\alpha_0 \,=\, 1/137.035999084,
\qquad\quad
\Delta\alpha^{(5)}_{\rm had}(M_Z)_0 = 0.027660,
\nonumber
\\
&&
G_{F 0} \,=\, 1.1663787 \times 10^{-5},
\qquad\quad
\alpha_{S0}^{(5)}(M_Z) = 0.1179,\phantom{xxx}
\nonumber
\\
&&
M_{t0} \,=\, \mbox{172.5 GeV},\qquad\quad M_{h0} \,=\, \mbox{125.25 GeV},\qquad\quad M_{Z0} \,=\, \mbox{91.1876 GeV},
\nonumber
\\
&&
m_b(m_b)_0 \,=\, \mbox{4.18 GeV},\qquad
m_c(m_c)_0 \,=\, \mbox{1.27 GeV},\qquad
m_s(\mbox{2 GeV})_0 \,=\, \mbox{93 MeV},
\nonumber
\\
&&
m_u(\mbox{2 GeV})_0 \,=\, \mbox{2.16 MeV},
\qquad
m_d(\mbox{2 GeV})_0 \,=\, \mbox{4.67 MeV},
\nonumber
\\
&&
M_{\tau 0} \,=\, \mbox{1.77686 GeV},
\qquad
M_{\mu 0} \,=\, \mbox{0.1056583745 GeV},
\nonumber
\\
&&
M_{e 0} \,=\, \mbox{0.5109989461 MeV}
.
\eeq
The Sommerfeld fine structure constant $\alpha$ is very accurately known compared to the others,
with a fractional uncertainty of $1.5 \times 10^{-10}$, so no variation in it will be considered.

For the other on-shell parameters, we next define the following dimensionless quantities:
\beq
\delta_Z &=& (M_Z - M_{Z0})/({\text{0.001 GeV}}),\\
\delta_t &=& (M_t - M_{t0})/({\text{1 GeV}}),\\
\delta_h &=& (M_h - M_{h0})/({\text{0.1 GeV}}),\\
\delta_S &=& 1000 \left [\alpha_S^{(5)}(M_Z) - \alpha_{S0}^{(5)}(M_Z) \right ],\\
\delta_a &=& 10^4 \left[ 
\Delta \alpha_{\rm had}^{(5)}(M_Z) - \Delta \alpha_{\rm had, 0}^{(5)}(M_Z)
\right ]
,
\eeq
as measures of the deviation from the benchmark model. The normalizations of these five quantities are chosen so that a change in the on-shell input by an amount of order the present experimental uncertainty will correspond very roughly to an order 1 change in the corresponding $\delta$. In the interpolation formulas found below, we will often give at least the contributions linear in the five $\delta$'s above, even when they are numerically too small to be practically significant, in order to quantitatively illustrate their contributions to the parametric errors. 

In the interpolation formulas for the Yukawa couplings at $Q=200$ GeV, we will also make use of the variations in the fermion masses as needed, parameterized by
\beq
\Delta_{f} \,=\, \frac{m_f}{m_{f0}} - 1,
\eeq
for $f=b,c,s,u,d,\tau,\mu,e$, where the $m_f$ are running \MSbar masses $m_b(m_b)$, $m_c(m_c)$, $m_s$(2 GeV), $m_u$(2 GeV), and $m_d$(2 GeV) for the light quarks, and pole (on-shell) masses for the leptons $m_f = M_\tau$, $M_\mu$, and $M_e$. Also, in a few of the interpolation formulas, we will include the small effect due to a possible deviation in the Fermi decay
constant from its benchmark central value, parameterized by
\beq
\Delta_{G_F} \,=\, \frac{G_F}{G_{F0}} - 1.
\eeq
The effects of this are typically expected to be very small, since the fractional uncertainty in $G_F$ given in the RPP is about $5 \times 10^{-7}$.

The interpolation formulas presented below were obtained by running v1.2 of {\tt SMDR} with its default choices repeatedly for points in parameter space on grids that cover the plausible allowed ranges, and then performing a fit to obtain the coefficients, which were then validated on more parameter space grids. We aim to provide results accurate to well under the experimental and theoretical uncertainties, for deviations of the on-shell inputs by up to 5 times their RPP quoted experimental uncertainties. Contributions quadratic in the deviations therefore will also be included when necessary to achieve relative precision goals for each quantity as stated below. For each output parameter, we will quote a conservative fractional precision, which in this paper refers to the fractional difference between the interpolation formula result and the output of {\tt SMDR} with default scale-setting choices, obtained as all on-shell inputs are varied over ranges such that the total deviation from the experimental central values, added in quadrature, is $\leq 5\sigma$. Here we interpret the uncertainties quoted in the RPP as $1\sigma$, even though a Gaussian distribution of errors may not be the appropriate description. It should be recognized that the actual theoretical uncertainty and the parametric uncertainty are both always much larger than this fractional precision. We have attempted to err on the side of including coefficients even when they are only significant for rather large deviations from the experimental central values.

We now provide the benchmark output results obtained using v1.2 of {\tt SMDR} with default choices. We give many more significant digits than justified by the theoretical and parametric uncertainties, merely for the sake of reproducibility. The benchmark running \MSbar parameters evaluated at the scale $Q = 200$ GeV are:
\beq
g_{30} &=& 
1.1525136966,
\label{eq:gthreebenchmark}
\\
g_0  &=& 
0.64683244428,
\label{eq:gbenchmark}
\\
g'_0 &=& 
0.35885152738,
\label{eq:gpbenchmark}
\\
\lambda_0  &=& 
0.12353343830,
\label{eq:lambdabenchmark}
\\
m^2_{0} &=& -(\mbox{%
93.126827678
GeV})^2,
\label{eq:m2benchmark}
\\
y_{t0} &=& 
0.92377763013,
\label{eq:ytbenchmark}
\\
y_{b0} &=& 
0.0153349059085,
\label{eq:ybbenchmark}
\\
y_{c0} &=& 
0.00336181598480,
\label{eq:ycbenchmark}
\\
y_{s0} &=& 
2.8885955612 \times 10^{-4},
\label{eq:ysbenchmark}
\\
y_{d0} &=& 
1.4505079604 \times 10^{-5},
\label{eq:ydbenchmark}
\\
y_{u0} &=& 
6.6738103560 \times 10^{-6},
\label{eq:yubenchmark}
\\
y_{\tau 0} &=& 
0.0100065524355,
\label{eq:ytaubenchmark}
\\
y_{\mu 0} &=& 
5.8908805223 \times 10^{-4},
\label{eq:ymubenchmark}
\\
y_{e 0} &=& 
2.7963423115 \times 10^{-6}
.
\label{eq:yebenchmark}
\eeq
Also, the physical $W$ boson mass in the PDG parameterization is found to be, for this benchmark set of parameters,
\beq
M_{W0} &=& \mbox{80.352476 GeV},
\label{eq:MWbenchmark}
\eeq
where we have used the {\tt SMDR} default by computing the $W$ boson pole mass in terms of the running parameters at $Q=160$ GeV. The values in eqs.~(\ref{eq:gthreebenchmark})-(\ref{eq:MWbenchmark}) will be used in the interpolation formulas below, as they give the results when all of the $\delta$'s vanish, by definition.

\section{Interpolation formula for the $W$-boson mass \label{sec:Wmass}}
\setcounter{equation}{0}
\setcounter{figure}{0}
\setcounter{table}{0}
\setcounter{footnote}{1}

For the $W$-boson physical mass in the PDG convention,\footnote{The result for $M_W$ has recently become of heightened interest because of a report \cite{CDF:2022hxs} from the Fermilab Tevatron's CDF collaboration which is incompatible with the Standard Model prediction, and in strong tension with other experimental results \cite{ParticleDataGroup:2022pth}.} we find
\beq
M_W &=& 
M_{W0} \, \bigl (1 
+ c^t_{M_W} \delta_t 
+ c^Z_{M_W} \delta_Z 
+ c^a_{M_W} \delta_a 
+ c^S_{M_W} \delta_S 
+ c^h_{M_W} \delta_h 
+ c^{tt}_{M_W} \delta_t^2 
\bigr )
,
\label{eq:Mwshort}
\eeq
where $M_{W0}$ was given in eq.~(\ref{eq:MWbenchmark}), and the other potentially significant coefficients are
\beq
&&
c^t_{M_W} = 7.61\times 10^{-5},\qquad
c^Z_{M_W} = 1.56\times 10^{-5}, \qquad
c^a_{M_W} = -2.29\times 10^{-5},\qquad
\nonumber
\\[2pt]
&&
c^S_{M_W} = -8.8\times 10^{-6},\qquad
c^h_{M_W} = -5.9\times 10^{-7},\qquad
c^{tt}_{M_W} = 1.3\times 10^{-7}.\phantom{xxxx}
\label{eq:MWcoeffs}
\eeq
This interpolation formula reproduces the results of {\tt SMDR} (with its default scale-setting choices) to better than 0.1 MeV, which is much smaller than the current theoretical and experimental uncertainties, when the input on-shell parameters are varied such that the total deviation from the central values, added in quadrature, is $\leq 5\sigma$. 

The results above are based on the pure \MSbar scheme used by {\tt SMDR}, and can be compared with similar interpolation formula results based on on-shell \cite{Awramik:2003rn} and hybrid \cite{Degrassi:2014sxa} scheme calculations, which both used fits to a much wider range for the Higgs mass. A numerical comparison between the results from these three different approaches was made in ref.~\cite{Martin:2022qiv} (see in particular Figures 4.1 and 4.2), showing that they agree well within the theoretical uncertainty due to renormalization scale dependence, and supporting a theoretical error estimate of perhaps $\pm 4$ MeV. This is less than the parametric error, coming principally from the top-quark mass, of about $6.11 \delta_t + 1.25 \delta_Z - 1.84 \delta_a - 0.71 \delta_S - 0.047 \delta_h + 0.010 \delta_t^2$, in MeV, which can be read off from eq.~(\ref{eq:MWcoeffs}).
The relatively large uncertainty associated with the top-quark mass is difficult to reduce, since it is due in large part to the problems in connecting hadron collider measurements and simulations to a well-defined short-distance top-quark mass or Yukawa coupling. 

\vspace{-6pt}

\section{Interpolation formulas for the \MSbar parameters \label{sec:MSbar}}
\setcounter{equation}{0}
\setcounter{figure}{0}
\setcounter{table}{0}
\setcounter{footnote}{1}

\vspace{-6pt}

\subsection{Higgs sector}

For the Higgs self-coupling $\lambda$ at $Q = 200$ GeV, we find
\beq
\lambda &=& 
\lambda_0 \, \bigl (1 
+ c^h_\lambda \delta_h 
+ c^t_\lambda \delta_t 
+ c^Z_\lambda \delta_Z 
+ c^S_\lambda \delta_S 
+ c^a_\lambda \delta_a 
+ c^{tt}_\lambda \delta_t^2 
+ c^{tS}_\lambda \delta_t \delta_S 
+ c^{hh}_{\lambda} \delta_h^2 
\nonumber
\\[1.5pt]
&& 
+ c^{ht}_\lambda \delta_h \delta_t 
+ c^{SS}_\lambda \delta_S^2 
+ c^{hS}_\lambda \delta_h \delta_S
+ c^{ttt}_\lambda\,\delta_t^3
+ c^{ttS}_\lambda\, \delta_t^2 \delta_S
+ c^{b}_\lambda\, \Delta_b
+ c^{G_F}_\lambda\, \Delta_{G_F}
\bigr )
,
\label{eq:lambda}
\eeq
where $\lambda_0$ was given in eq.~(\ref{eq:lambdabenchmark}), and the other coefficients are
\beq
&&
c^h_\lambda   =  1.6823\times 10^{-3}, \qquad
c^t_\lambda   =  -1.488\times 10^{-4}, \qquad
c^Z_\lambda   =  -3.5\times 10^{-7}, \qquad
\nonumber
\\[1pt]
&&
c^S_\lambda   =  -2.2\times 10^{-7}, \qquad
c^a_\lambda   =  3.4\times 10^{-7}, \qquad
c^{tt}_\lambda   =  1.528\times 10^{-5}, \qquad
\nonumber
\\[1pt]
&&
c^{tS}_\lambda   =  -4.02\times 10^{-6},  \qquad
c^{hh}_\lambda   =  7.0\times 10^{-7}, \qquad
c^{ht}_\lambda  = -6.1\times 10^{-7},  \qquad
\nonumber
\\[1pt]
&&
c^{SS}_\lambda  =  3.0\times 10^{-7}, \qquad
c^{hS}_\lambda   = 6.4\times 10^{-8},\qquad
c^{ttt}_\lambda   =  1.9\times 10^{-7}, \qquad
\nonumber
\\[1pt]
&&
c^{ttS}_\lambda   =  -7.6\times 10^{-8},\qquad
c^{b}_\lambda = 4.5 \times 10^{-5},\qquad
c^{G_F}_\lambda = 0.95
.
\label{eq:lambdacoeffs}
\eeq
This formula, based on a fit to the best available calculation of the physical Higgs boson mass \cite{Martin:2014cxa,Martin:2022qiv}, agrees with the results of {\tt SMDR} to better than $10^{-6}$ fractional precision in $\lambda$ as the input parameters are varied over ranges with a total deviation, added in quadrature, of $5\sigma$ from their central values. Again, the theoretical and parametric errors are much larger than this fractional precision, with the top-quark mass giving the largest contribution to the error budget other than the Higgs boson mass itself.

For the running Higgs squared mass parameter at $Q=200$ GeV, we find
\beq
m^2 &=& 
m^2_0 \,\bigl (1 
+ c^h_{m^2} \delta_h 
+ c^t_{m^2} \delta_t 
+ c^S_{m^2} \delta_S 
+ c^Z_{m^2} \delta_Z 
+ c^a_{m^2} \delta_a 
+ c^{tt}_{m^2} \delta_t^2 
+ c^{tS}_{m^2} \delta_t \delta_S 
\nonumber
\\[1.5pt]
&&
+ c^{hh}_{m^2} \delta_h^2
+ c^{ht}_{m^2} \delta_h \delta_t 
\bigr )
,
\label{eq:m2}
\eeq
where $m_0^2$ was given in eq.~(\ref{eq:m2benchmark}), and the other 
significant coefficients are
\beq
&&
c^h_{m^2} =  1.4319\times 10^{-3}, \qquad\,
c^t_{m^2} =  2.337\times 10^{-3}, \qquad\,
c^S_{m^2} =  -1.052\times 10^{-4}, 
\nonumber
\\[1pt]
&&
c^Z_{m^2} =  -5.7\times 10^{-7}, \qquad\>\>\,\,
c^a_{m^2} = 5.4\times 10^{-7}, \qquad\>\>\>\>
c^{tt}_{m^2} = 2.02\times 10^{-5}, 
\nonumber
\\[1pt]
&&
c^{tS}_{m^2} = -2.45\times 10^{-6}, \qquad\>\>
c^{hh}_{m^2} = 5.8\times 10^{-7}, \qquad\>\>\>
c^{ht}_{m^2} = -4.3\times 10^{-7}
.
\label{eq:m2coeffs}
\eeq
This formula provides agreement with the output of {\tt SMDR} to a fractional precision of better than $10^{-5}$.

\vspace{-6pt}

\subsection{Gauge couplings}

For the $SU(3)_c$ \MSbar gauge coupling $g_3$ evaluated at $Q=200$ GeV, we obtained the following interpolation formula:
\beq
g_3 \,=\, 
g_{30} \, \bigl (1 
+ c_{g_3}^S \delta_S 
+ c_{g_3}^t \delta_t 
+ c_{g_3}^{SS} \delta_S^2
+ c_{g_3}^h \delta_h 
+ c_{g_3}^Z \delta_Z 
+ c_{g_3}^a \delta_a 
\bigr )
,
\label{eq:g3}
\eeq
where $g_{30}$ was given in eq.~(\ref{eq:gthreebenchmark}), and the coefficients are
\beq
&&
c_{g_3}^S \,=\,  3.7875\times 10^{-3}, \qquad
c_{g_3}^t \,=\,   -3.98\times 10^{-5}, \qquad
c_{g_3}^{SS} \,=\, -1.07\times 10^{-5}, \qquad
\nonumber \\[1pt] &&
c_{g_3}^h \,=\,   2.5\times 10^{-8}, \qquad\>\>\,
c_{g_3}^Z \,=\,   2.7\times 10^{-9}, \qquad\>\,\,
c_{g_3}^a \,=\,  -2.0\times 10^{-9}
.
\label{eq:g3coeffs}
\eeq
Note that the top-quark mass is significant here because we are relating the 5-quark QCD coupling $\alpha_S^{(5)}(M_Z)$ to the Standard Model QCD coupling 
$g_3$ with the top quark not decoupled.
The three terms proportional to $\delta_S$, $\delta_t$, and $\delta_S^2$ are sufficient to obtain a fractional precision compared to {\tt SMDR} of better than $10^{-5}$, but the linear deviation coefficients $c_{g_3}^h$, $c_{g_3}^Z$, and $c_{g_3}^a$
are also listed in order to illustrate the small size of the parametric errors.

For the $SU(2)_L$ gauge coupling $g$, we find:
\beq
g &=& 
g_0\, \bigl ( 1 +c^t_{g} \delta_t + c^a_g \delta_a + c^Z_g \delta_Z  + c^S_g \delta_S + c^h_g \delta_h + c^{tt}_g \delta_t^2 + c^{tS}_g \delta_t \delta_S + 
c_g^{G_F} \Delta_{G_F}\bigr )
,
\label{eq:gshort}
\eeq
where $g_0$ was given in eq.~(\ref{eq:gbenchmark}), and the other coefficients are
\beq
&&
c^t_g   \,=\,  5.735\times 10^{-5}, \qquad\>\>
c^a_g   \,=\,  -2.295\times 10^{-5}, \qquad
c^Z_g   \,=\,  1.558\times 10^{-5}, \qquad
\nonumber
\\[1pt]
&&
c^S_g   \,=\,  -5.97\times 10^{-6}, \qquad
c^h_g   \,=\,  -8.5\times 10^{-7}, \qquad\>\>\>\>
c^{tt}_g  \,=\,  1.9\times 10^{-7}, \qquad
\nonumber
\\[1pt]
&&
c^{tS}_g  \,=\,  -7.8\times 10^{-8},
\qquad\>
c^{G_F}_g  \,=\,  0.71.
\label{eq:gcoeffs}
\eeq
This formula provides a fractional precision of better than $10^{-6}$ as the input on-shell
parameters are varied with $\leq 5 \sigma$ total deviation from their central values, added in quadrature.

For the $U(1)_Y$ gauge coupling, we find
\beq
g' = 
&&
g'_0 \,\bigl 
(1 + c^t_{g'} \delta_t + c^a_{g'} \delta_a + c^Z_{g'} \delta_Z  + c^S_{g'} \delta_S + c^h_{g'} \delta_h
\bigr )
,
\label{eq:gp}
\eeq
where $g'_0$ was given in eq.~(\ref{eq:gpbenchmark}) and the other coefficients are
\beq
&&
c^t_{g'}   \,=\,  -2.609\times 10^{-5}, \qquad
c^a_{g'}   \,=\,  7.714\times 10^{-5}, \qquad
c^Z_{g'}   \,=\,  -4.70\times 10^{-6}, \qquad
\nonumber
\\
&&
c^S_{g'}   \,=\,  3.29\times 10^{-6}, \qquad
c^h_{g'}   \,=\,  2.6\times 10^{-7}. \qquad
\label{eq:gpcoeffs}
\eeq
This formula again provides a fractional precision of better than $10^{-6}$ compared to {\tt SMDR}.

\vspace{-6pt}

\subsection{Top-quark Yukawa coupling}

For the top-quark Yukawa coupling at $Q = 200$ GeV, we find
\beq
y_t &=& 
y_{t0}\, \bigl (1 
+ c^t_{y_t} \delta_t 
+ c^S_{y_t} \delta_S 
+ c^h_{y_t} \delta_h 
+ c^{tt}_{y_t} \delta_t^2 
+ c^{SS}_{y_t} \delta_S^2
+ c^Z_{y_t} \delta_Z 
+ c^a_{y_t} \delta_a 
\bigr )
,
\label{eq:yt}
\eeq
where $y_{t0}$ was given in eq.~(\ref{eq:ytbenchmark}), and the other coefficients are
\beq
&&
c^t_{y_t}   =  6.352\times 10^{-3}, \qquad
c^S_{y_t}   =  -7.76\times 10^{-4}, \qquad
c^h_{y_t}   =  -2.36\times 10^{-6}, \qquad
\nonumber
\\[1pt]
&&
c^{tt}_{y_t}   =  8.9\times 10^{-7}, \qquad\>\>\>\,
c^{SS}_{y_t}  =  -1.23\times 10^{-6}, \qquad\>
c^Z_{y_t}   =  -1.6\times 10^{-7}, \qquad
\nonumber
\\[1pt]
&&
c^a_{y_t}   =  2.2\times 10^{-8}
.
\label{eq:ytcoeffs}
\eeq
The five terms proportional to $\delta_t$, $\delta_S$, $\delta_h$, $\delta_t^2$, and $\delta_S^2$ are sufficient to obtain a fractional precision better than $10^{-5}$, and the linear deviation coefficients $c_{y_t}^Z$ and $c_{y_t}^a$ are also included in order to show their small contribution to the parametric error budget.

\vspace{-6pt}

\subsection{Yukawa couplings of light quarks}

In the interpolation formulas for light-quark Yukawa couplings in the present subsection, the quantities $\delta_a$, $\delta_h$, 
and $\delta_Z$ make a relatively insignificant difference, and are therefore omitted.

For the bottom-quark Yukawa coupling at $Q=200$ GeV, we find
\beq
y_b &=& y_{b0}\,\bigl ( 
1 
+ c^b_{y_b} \Delta_{b} 
+ c^{bb}_{y_b} \Delta_{b}^2 
+ c^{bS}_{y_b} \Delta_{b} \delta_S
+ c^S_{y_b} \delta_S
+ c^t_{y_b} \delta_t
+ c^{SS}_{y_b} \delta_S^2
+ c^{SSS}_{y_b} \delta_S^3
\bigr ) ,
\eeq
where $y_{b0}$ was given in eq.~(\ref{eq:ybbenchmark}), and the other coefficients  are
\beq
&&
c_{y_b}^b = 1.185,\qquad 
c_{y_b}^{bb} = 0.075,\qquad 
c_{y_b}^{bS} = -3.3\times 10^{-3},\qquad
c^S_{y_b} = -6.125 \times 10^{-3},\qquad
\nonumber
\\[1pt]
&&
c_{y_b}^{t} = -2.4 \times 10^{-5}
,\qquad
c_{y_b}^{SS} = -2.1\times 10^{-5}
,\qquad
c_{y_b}^{SSS} = -1.5 \times 10^{-7}
.
\eeq
This agrees with the results of {\tt SMDR} to a fractional precision of 
better than $10^{-4}$.
For the charm-quark Yukawa coupling at $Q=200$ GeV, we obtain 
\beq
y_c &=& y_{c0}\,\bigl ( 
1 
+ c^c_{y_c} \Delta_{c} 
+ c^{cc}_{y_c} \Delta_{c}^2 
+ c^{cS}_{y_c} \Delta_{c} \delta_S
+ c^S_{y_c} \delta_S
+ c^{SS}_{y_c} \delta_S^2
+ c^{SSS}_{y_c} \delta_S^3
\nonumber \\ &&
+ c^b_{y_c} \Delta_{b}
+ c^{bS}_{y_c} \Delta_{b} \delta_S
+ c^t_{y_c} \delta_{t}
\bigr ) ,\phantom{xxxx}
\eeq
where $y_{c0}$ was given in eq.~(\ref{eq:ycbenchmark}), and the other coefficients are
\beq
&&
c_{y_c}^c = 1.415,\qquad\>\>\> 
c_{y_c}^{cc} = 0.078,\qquad\>\>\>
c_{y_c}^{cS} = -3.0\times 10^{-3},\qquad
\nonumber
\\[1pt]
&&
c_{y_c}^S = -0.01746,\qquad
c_{y_c}^{SS} = -2.34\times 10^{-4},\qquad
c_{y_c}^{SSS} = -6.5\times 10^{-6},\qquad
\nonumber
\\[1pt]
&&
c_{y_c}^b = -0.027,\qquad
c_{y_c}^{bS} = -1.6 \times 10^{-3}
,\qquad
c_{y_c}^{t} = -1.5 \times 10^{-5}.
\eeq
This result for the charm-quark Yukawa coupling agrees with {\tt SMDR} to a fractional precision of better than $10^{-4}$.

For the strange, down, and up Yukawa couplings, the interpolation
formulas have a simpler, universal form, due to the fact that the ``on-shell" input parameters
from the RPP are actually running \MSbar parameters determined 
at a common scale of $Q=2$ GeV, so that the same QCD corrections apply to all three in the same way. For the Yukawa couplings at $Q=200$ GeV, we find:
\beq
y_q &=& y_{q0}\, 
\bigl (1 + \Delta_{q} \bigr ) \left (1 
+ c^S_{y_q} \delta_S 
+ c^{SS}_{y_q} \delta_S^2
+ c^{SSS}_{y_q} \delta_S^3
+ c^b_{y_q} \Delta_b
+ c^t_{y_q} \delta_t
\right )
,
\eeq
where  the coefficients in all three cases $(q=s,d,u)$ are approximated well by
\beq
&&
c^S_{y_q} = -0.01089,\qquad
c^{SS}_{y_q} = -7.93 \times 10^{-5},\qquad
c^{SSS}_{y_q} = -1.2 \times 10^{-6},
\nonumber \\ 
&&
c^b_{y_q} = -0.0128,\qquad
c_{y_q}^{t} = -1.5 \times 10^{-5},
\eeq
and $y_{s 0}$, $y_{d0}$, and $y_{u0}$ were given respectively in eqs.~(\ref{eq:ysbenchmark}), (\ref{eq:ydbenchmark}), and (\ref{eq:yubenchmark}).
These formulas agree with those obtained by {\tt SMDR} to a fractional precision of better than
$10^{-4}$.

\vspace{-6pt}

\subsection{Yukawa couplings of leptons}

For the tau-lepton Yukawa coupling at $Q=200$ GeV, we obtain
\beq
y_\tau &=& y_{\tau 0}\,\bigl ( 
1 + \Delta_{\tau} + 0.5 \Delta_{G_F}
+ c^t_{y_\tau} \delta_t
+ c^S_{y_\tau} \delta_S
+ c^a_{y_\tau} \delta_a
+ c^h_{y_\tau} \delta_h
+ c^Z_{y_\tau} \delta_Z
\nonumber \\[4pt] &&
+ c^{tt}_{y_\tau} \delta_t^2
+ c^{tS}_{y_\tau} \delta_t \delta_S
\bigr )
,
\phantom{xxx}
\eeq
where $y_{\tau 0}$ was given in eq.~(\ref{eq:ytaubenchmark}), and the coefficients of $\Delta_{\tau}$ and $\Delta_{G_F}$ are very close to 1 and 0.5 as indicated, and the other
coefficients are
\beq
&& 
c^t_{y_\tau} = -1.252 \times 10^{-5}, \qquad
c^S_{y_\tau} =  2.63 \times 10^{-6}, \qquad
c^a_{y_\tau} = -1.83 \times 10^{-6}, \qquad
\nonumber \\[1pt]
&&
c^h_{y_\tau} =  1.74 \times 10^{-6}, \qquad
c^Z_{y_\tau} = -1.8 \times 10^{-7}, \qquad
c^{tt}_{y_\tau} =  -6.9 \times 10^{-7}, \qquad
\nonumber \\[1pt] 
&&
c^{tS}_{y_\tau} =  1.3 \times 10^{-7}. \qquad
\eeq
This interpolation formula gives agreement with {\tt SMDR} to a fractional precision of
better than $10^{-7}$.

The Yukawa couplings for $\ell = \mu, e$ at $Q= 200$ GeV are written in the common form:
\beq
y_\ell &=& y_{\ell 0}\,\bigl ( 
1 + \Delta_{\ell} + 0.5 \Delta_{G_F}
+ c^t_{y_\ell} \delta_t
+ c^S_{y_\ell} \delta_S
+ c^a_{y_\ell} \delta_a
+ c^h_{y_\ell} \delta_h
+ c^Z_{y_\ell} \delta_Z
\nonumber \\[3pt] &&
+ c^{tt}_{y_\ell} \delta_t^2
+ c^{tS}_{y_\ell} \delta_t \delta_S
+ c^{c}_{y_\ell} \Delta_{c}
+ c^{b}_{y_\ell} \Delta_{b}
\bigr )
,
\phantom{xxx}
\eeq
where $y_{\mu 0}$ and $y_{e 0}$ were given in eqs.~(\ref{eq:ymubenchmark}) and (\ref{eq:yebenchmark}), respectively.
For the muon, the other coefficients are:
\beq
&& 
c^t_{y_\mu} = -1.3105 \times 10^{-5}, \qquad
c^S_{y_\mu} =  2.17 \times 10^{-6}, \qquad
c^a_{y_\mu} = -2.84 \times 10^{-6}, \qquad
\nonumber \\[1pt]
&&
c^h_{y_\mu} =  1.73 \times 10^{-6}, \qquad
c^Z_{y_\mu} = -1.78 \times 10^{-7}, \qquad
c^{tt}_{y_\mu} =  -6.93 \times 10^{-7}, \qquad
\nonumber \\[1pt] 
&&
c^{tS}_{y_\mu} =  1.26 \times 10^{-7}, \qquad
c^{c}_{y_\mu} = -3.3 \times 10^{-5}, \qquad
c^{b}_{y_\mu} = -4.1 \times 10^{-6}.
\eeq
For the electron, the coefficients are
\beq
&& 
c^t_{y_e} = -1.312 \times 10^{-5}, \qquad
c^S_{y_e} =  2.87 \times 10^{-6}, \qquad
c^a_{y_e} = -4.72 \times 10^{-6}, \qquad
\nonumber \\[1pt] 
&&
c^h_{y_e} =  1.73 \times 10^{-6}, \qquad
c^Z_{y_e} = -1.78 \times 10^{-7}, \qquad
c^{tt}_{y_e} =  -6.93 \times 10^{-7}, \qquad
\nonumber \\[1pt] 
&&
c^{tS}_{y_e} =  1.26 \times 10^{-7}, \qquad
c^{c}_{y_e} = -8.1 \times 10^{-5}, \qquad
c^{b}_{y_e} = -1.4 \times 10^{-5}.
\eeq
The fractional precisions, compared to the results from {\tt SMDR}, are less than $10^{-9}$.
Since the present fractional uncertainties in $M_\mu$ and $M_e$ are about
$2 \times 10^{-8}$ and $6 \times 10^{-9}$  respectively, we see that for each lepton,
the bottleneck for obtaining the most accurate possible Yukawa coupling in the ultraviolet is not the uncertainty in the corresponding lepton mass, but rather the uncertainty associated with the top-quark mass, which is difficult to reduce as we have already mentioned.

\vspace{-10pt}

\section{Outlook \label{sec:outlook}}
\setcounter{equation}{0}
\setcounter{figure}{0}
\setcounter{table}{0}
\setcounter{footnote}{1}

In this paper we have presented simple interpolation formulas that provide the 
fundamental Lagrangian parameters for the Standard Model, given the corresponding on-shell experimental values as inputs. (The three physical angles and CP-violating phase associated with CKM mixing are omitted, having a tiny effect on these results due to CKM unitarity,
and can be obtained from ref.~\cite{ParticleDataGroup:2022pth} and sources referenced therein.) These results are an alternative to a more time-consuming and
complicated evaluation using e.g.~the computer code {\tt SMDR}, on which our results are based. The structure of the interpolation formulas has been designed so as to avoid any numerically significant loss of precision, and are made to provide results at the \MSbar renormalization scale $Q=200$ GeV as a reference. For convenience, we have included as an ancillary file with this paper a simple interactive command-line Python code {\tt sm200.py} implementing the interpolation formulas above. We intend to update our results in the preprint version of this paper and in that code as new theoretical refinements and experimental measurements become available.

Besides satisfying  basic curiosity about the fundamental parameters of the Standard Model, the results given here will have applications in matching to various candidate ultraviolet completions of the Standard Model, provided that the mass scales associated with new physics are sufficiently high that non-renormalizable terms in the effective theory can be neglected or corrected for. The results also can be viewed as providing the parametric error budget for the defining couplings of the Standard Model Lagrangian.

\vspace{3pt}

{\it Acknowledgments:} 
This work was supported in part by the National Science Foundation grant number 
2013340.

\pagebreak



\begin{thebibliography}{90}
\baselineskip=14.85pt

\bibitem{Bardeen:1978yd}
  W.~A.~Bardeen, A.~J.~Buras, D.~W.~Duke and T.~Muta,
  ``Deep Inelastic Scattering Beyond the Leading Order in
  Asymptotically Free Gauge Theories,''
  Phys.\ Rev.\ D {\bf 18}, 3998 (1978).

\bibitem{Braaten:1981dv}
  E.~Braaten and J.~P.~Leveille,
  ``Minimal Subtraction and Momentum Subtraction in {QCD} at Two Loop Order,''
  Phys.\ Rev.\ D {\bf 24}, 1369 (1981).

\bibitem{Bollini:1972ui}
  C.~G.~Bollini and J.~J.~Giambiagi,
  ``Dimensional Renormalization: The Number of Dimensions as a
  Regularizing Parameter,''
  Nuovo Cim.\ B {\bf 12}, 20 (1972).
 C.~G.~Bollini and J.~J.~Giambiagi,
  ``Lowest order divergent graphs in nu-dimensional space,''
  Phys.\ Lett.\ B {\bf 40}, 566 (1972).

\bibitem{Ashmore:1972uj}
  J.~F.~Ashmore,
  ``A Method of Gauge Invariant Regularization,''
  Lett.\ Nuovo Cim.\  {\bf 4}, 289 (1972).

\bibitem{Cicuta:1972jf}
  G.~M.~Cicuta and E.~Montaldi,
  ``Analytic renormalization via continuous space dimension,''
  Lett.\ Nuovo Cim.\  {\bf 4}, 329 (1972).

\bibitem{tHooft:1972fi}
  G.~'t Hooft and M.~J.~G.~Veltman,
  ``Regularization and Renormalization of Gauge Fields,''
  Nucl.\ Phys.\ B {\bf 44}, 189 (1972).

\bibitem{tHooft:1973mm}
  G.~'t Hooft,
  ``Dimensional regularization and the renormalization group,''
  Nucl.\ Phys.\ B {\bf 61}, 455 (1973).




\bibitem{MVI}
  M.~E.~Machacek and M.~T.~Vaughn,
  ``Two Loop Renormalization Group Equations in a General Quantum Field Theory.
  1. Wave Function Renormalization,''
  Nucl.\ Phys.\ B {\bf 222}, 83 (1983).

\bibitem{MVII}
  M.~E.~Machacek and M.~T.~Vaughn,
  ``Two Loop Renormalization Group Equations in a General Quantum Field Theory.
  2. Yukawa Couplings,''
  Nucl.\ Phys.\ B {\bf 236}, 221 (1984).

\bibitem{Jack:1984vj}
  I.~Jack and H.~Osborn,
  ``General Background Field Calculations With Fermion Fields,''
  Nucl.\ Phys.\ B {\bf 249}, 472 (1985).

\bibitem{MVIII}
  M.~E.~Machacek and M.~T.~Vaughn,
  ``Two Loop Renormalization Group Equations in a General Quantum Field Theory.
  3. Scalar Quartic Couplings,''
  Nucl.\ Phys.\ B {\bf 249}, 70 (1985).

\bibitem{Luo:2002ey}
  M.~x.~Luo and Y.~Xiao,
  ``Two loop renormalization group equations in the standard model,''
  Phys.\ Rev.\ Lett.\  {\bf 90}, 011601 (2003)
  [hep-ph/0207271].

\bibitem{Tarasov} O.V.~Tarasov,
``Anomalous Dimensions Of Quark Masses In Three Loop
Approximation,'' preprint JINR-P2-82-900, (1982), unpublished. (In Russian.)

\bibitem{Mihaila:2012fm}
  L.~N.~Mihaila, J.~Salomon and M.~Steinhauser,
  ``Gauge Coupling Beta Functions in the Standard Model to Three Loops,''
  Phys.\ Rev.\ Lett.\  {\bf 108}, 151602 (2012)
  [arXiv:1201.5868 [hep-ph]].

\bibitem{Chetyrkin:2012rz}
  K.~G.~Chetyrkin and M.~F.~Zoller,
  ``Three-loop $\beta$-functions for top-Yukawa and the Higgs
  self-interaction in the Standard Model,''
  JHEP {\bf 1206}, 033 (2012)
  [1205.2892].

\bibitem{Bednyakov:2012rb}
  A.~V.~Bednyakov, A.~F.~Pikelner and V.~N.~Velizhanin,
  ``Anomalous dimensions of gauge fields and gauge coupling beta-functions 
  in the Standard Model at three loops,''
  JHEP {\bf 1301}, 017 (2013)
  [arXiv:1210.6873 [hep-ph]].

\bibitem{Bednyakov:2012en}
  A.~V.~Bednyakov, A.~F.~Pikelner and V.~N.~Velizhanin,
  ``Yukawa coupling beta-functions in the Standard Model at three loops,''
  Phys.\ Lett.\ B {\bf 722}, 336 (2013)
  [arXiv:1212.6829 [hep-ph]].

\bibitem{Chetyrkin:2013wya}
  K.~G.~Chetyrkin and M.~F.~Zoller,
  ``$\beta$-function for the Higgs self-interaction in the
  Standard Model at three-loop level,''
  JHEP {\bf 1304}, 091 (2013)
  [1303.2890].

\bibitem{Bednyakov:2013eba}
  A.~V.~Bednyakov, A.~F.~Pikelner and V.~N.~Velizhanin,
  ``Higgs self-coupling beta-function in the Standard Model at three loops,''
  Nucl.\ Phys.\ B {\bf 875}, 552 (2013)
  [1303.4364].

\bibitem{Bednyakov:2013cpa}
  A.~V.~Bednyakov, A.~F.~Pikelner and V.~N.~Velizhanin,
  ``Three-loop Higgs self-coupling beta-function in the Standard Model with complex Yukawa matrices,''
  Nucl.\ Phys.\ B {\bf 879}, 256 (2014)
  [arXiv:1310.3806 [hep-ph]].

\bibitem{Bednyakov:2014pia}
  A.~V.~Bednyakov, A.~F.~Pikelner and V.~N.~Velizhanin,
  ``Three-loop SM beta-functions for matrix Yukawa couplings,''
  Phys.\ Lett.\ B {\bf 737}, 129 (2014)
  [arXiv:1406.7171 [hep-ph]].

\bibitem{vanRitbergen:1997va}
  T.~van Ritbergen, J.~A.~M.~Vermaseren and S.~A.~Larin,
  ``The Four loop beta function in quantum chromodynamics,''
  Phys.\ Lett.\ B {\bf 400}, 379 (1997)
  [hep-ph/9701390].

\bibitem{Czakon:2004bu}
  M.~Czakon,
  ``The Four-loop QCD beta-function and anomalous dimensions,''
  Nucl.\ Phys.\ B {\bf 710}, 485 (2005)
  [hep-ph/0411261].

\bibitem{Bednyakov:2015ooa} 
  A.~V.~Bednyakov and A.~F.~Pikelner,
  ``Four-loop strong coupling beta-function in the Standard Model,''
  Phys.\ Lett.\ B {\bf 762}, 151 (2016)
  [arXiv:1508.02680 [hep-ph]];
``On the four-loop strong coupling beta-function in the SM,''
  EPJ Web Conf.\  {\bf 125}, 04008 (2016)
  [arXiv:1609.02597 [hep-ph]].
  
\bibitem{Zoller:2015tha} 
  M.~F.~Zoller,
  ``Top-Yukawa effects on the $\beta$-function of the strong coupling in the SM at four-loop level,''
  JHEP {\bf 1602}, 095 (2016)
  [arXiv:1508.03624 [hep-ph]].

\bibitem{Poole:2019txl} 
  C.~Poole and A.~E.~Thomsen,
  ``Weyl Consistency Conditions and $\gamma_{5}$,''
  arXiv:1901.02749 [hep-th].

\bibitem{Baikov:2016tgj} 
  P.~A.~Baikov, K.~G.~Chetyrkin and J.~H.~Kuhn,
  ``Five-Loop Running of the QCD coupling constant,''
  Phys.\ Rev.\ Lett.\  {\bf 118}, no. 8, 082002 (2017)
  [arXiv:1606.08659 [hep-ph]].

\bibitem{Herzog:2017ohr} 
  F.~Herzog, B.~Ruijl, T.~Ueda, J.~A.~M.~Vermaseren and A.~Vogt,
  ``The five-loop beta function of Yang-Mills theory with fermions,''
  JHEP {\bf 1702}, 090 (2017)
  [arXiv:1701.01404 [hep-ph]].

\bibitem{Chetyrkin:1997dh}
  K.~G.~Chetyrkin,
  ``Quark mass anomalous dimension to ${\cal O}(\alpha_S^4$),''
  Phys.\ Lett.\ B {\bf 404}, 161 (1997)
  [hep-ph/9703278].

\bibitem{Vermaseren:1997fq}
  J.~A.~M.~Vermaseren, S.~A.~Larin and T.~van Ritbergen,
  ``The four loop quark mass anomalous dimension and the invariant quark mass,''
  Phys.\ Lett.\ B {\bf 405}, 327 (1997)
  [hep-ph/9703284].

\bibitem{Baikov:2014qja}
  P.~A.~Baikov, K.~G.~Chetyrkin and J.~H.~K\"uhn,
  ``Quark Mass and Field Anomalous Dimensions to ${\cal O}(\alpha_S^5)$,''
  JHEP {\bf 1410}, 076 (2014)
  [arXiv:1402.6611 [hep-ph]].

\bibitem{Martin:2015eia} 
  S.~P.~Martin,
  ``Four-loop Standard Model effective potential at leading order in QCD,''
  Phys.\ Rev.\ D {\bf 92}, no. 5, 054029 (2015)
  [arXiv:1508.00912 [hep-ph]].

\bibitem{Chetyrkin:2016ruf} 
  K.~G.~Chetyrkin and M.~F.~Zoller,
  ``Leading QCD-induced four-loop contributions to the $\beta$-function of 
  the Higgs self-coupling in the SM and vacuum stability,''
  JHEP {\bf 1606}, 175 (2016)
  [arXiv:1604.00853 [hep-ph]].

\bibitem{Davies:2019onf}
J.~Davies, F.~Herren, C.~Poole, M.~Steinhauser and A.~E.~Thomsen,
``Gauge Coupling $\beta$ Functions to Four-Loop Order in the Standard Model,''
Phys. Rev. Lett. \textbf{124}, no.7, 071803 (2020)
[arXiv:1912.07624 [hep-ph]].



\bibitem{Ford:1992pn}
  C.~Ford, I.~Jack and D.R.T.~Jones,
  ``The Standard model effective potential at two loops,''
  Nucl.\ Phys.\ B {\bf 387}, 373 (1992)
  [Erratum-ibid.\ B {\bf 504}, 551 (1997)]
  [hep-ph/0111190].

\bibitem{Martin:2001vx}
  S.~P.~Martin,
  ``Two loop effective potential for a general renormalizable theory 
  and softly broken supersymmetry,''
  Phys.\ Rev.\ D {\bf 65}, 116003 (2002)
  [hep-ph/0111209].

\bibitem{Martin:2013gka}
  S.~P.~Martin,
  ``Three-loop Standard Model effective potential at leading order 
  in strong and top Yukawa couplings,''
  Phys.\ Rev.\ D {\bf 89}, no. 1, 013003 (2014)
  [1310.7553].

\bibitem{Martin:2016bgz}
S.~P.~Martin and D.~G.~Robertson,
``Evaluation of the general 3-loop vacuum Feynman integral,''
Phys. Rev. D \textbf{95}, no.1, 016008 (2017)
doi:10.1103/PhysRevD.95.016008
[arXiv:1610.07720 [hep-ph]].

\bibitem{Martin:2017lqn} 
  S.~P.~Martin,
  ``Effective potential at three loops,''
  Phys.\ Rev.\ D {\bf 96}, no. 9, 096005 (2017)
  [arXiv:1709.02397 [hep-ph]].
  

\bibitem{ParticleDataGroup:2022pth}
R.~L.~Workman \textit{et al.} [Particle Data Group],
``Review of Particle Physics,''
PTEP \textbf{2022}, 083C01 (2022)
doi:10.1093/ptep/ptac097


\bibitem{Willenbrock:2022smq}
S.~Willenbrock,
``Mass and width of an unstable particle,''
[arXiv:2203.11056 [hep-ph]].

\bibitem{Weinberg:1980wa} 
  S.~Weinberg,
  ``Effective Gauge Theories,''
  Phys.\ Lett.\  {\bf 91B}, 51 (1980).
  doi:10.1016/0370-2693(80)90660-7

\bibitem{Ovrut:1980dg} 
  B.~A.~Ovrut and H.~J.~Schnitzer,
  ``The Decoupling Theorem and Minimal Subtraction,''
  Phys.\ Lett.\  {\bf 100B}, 403 (1981).
  doi:10.1016/0370-2693(81)90146-5

\bibitem{Tarrach:1980up} 
  R.~Tarrach,
  ``The Pole Mass in Perturbative QCD,''
  Nucl.\ Phys.\ B {\bf 183}, 384 (1981).
  doi:10.1016/0550-3213(81)90140-1

\bibitem{Bernreuther:1981sg} 
  W.~Bernreuther and W.~Wetzel,
  ``Decoupling of Heavy Quarks in the Minimal Subtraction Scheme,''
  Nucl.\ Phys.\ B {\bf 197}, 228 (1982)
  Erratum: [Nucl.\ Phys.\ B {\bf 513}, 758 (1998)].
  doi:10.1016/0550-3213(82)90288-7, 10.1016/S0550-3213(97)00811-0

\bibitem{Bohm:1986rj}
  M.~Bohm, H.~Spiesberger and W.~Hollik,
  ``On the One Loop Renormalization of the Electroweak Standard Model
  and Its Application to Leptonic Processes,''
  Fortsch.\ Phys.\  {\bf 34}, 687 (1986).

\bibitem{vanderBij:1986hy} 
  J.~J.~van der Bij and F.~Hoogeveen,
  ``Two Loop Correction to Weak Interaction Parameters Due to a Heavy Fermion Doublet,''
  Nucl.\ Phys.\ B {\bf 283}, 477 (1987).

\bibitem{Djouadi:1987gn} 
  A.~Djouadi and C.~Verzegnassi,
  ``Virtual Very Heavy Top Effects in LEP / SLC Precision Measurements,''
  Phys.\ Lett.\ B {\bf 195}, 265 (1987).
  doi:10.1016/0370-2693(87)91206-8

\bibitem{Gray:1990yh} 
  N.~Gray, D.~J.~Broadhurst, W.~Grafe and K.~Schilcher,
  ``Three Loop Relation of Quark (Modified) Ms and Pole Masses,''
  Z.\ Phys.\ C {\bf 48}, 673 (1990).

\bibitem{Kniehl:1989yc} 
  B.~A.~Kniehl,
  ``Two Loop Corrections to the Vacuum Polarizations in Perturbative QCD,''
  Nucl.\ Phys.\ B {\bf 347}, 86 (1990).
  doi:10.1016/0550-3213(90)90552-O

\bibitem{Halzen:1990je} 
  F.~Halzen and B.~A.~Kniehl,
  ``$\Delta$ r beyond one loop,''
  Nucl.\ Phys.\ B {\bf 353}, 567 (1991).
  doi:10.1016/0550-3213(91)90319-S

\bibitem{Arason:1991ic}
H.~Arason, D.~J.~Castano, B.~Keszthelyi, S.~Mikaelian, E.~J.~Piard, P.~Ramond and B.~D.~Wright,
``Renormalization group study of the standard model and its extensions. 1. The Standard model,''
Phys. Rev. D \textbf{46}, 3945-3965 (1992)
doi:10.1103/PhysRevD.46.3945

\bibitem{Barbieri:1992nz} 
  R.~Barbieri, M.~Beccaria, P.~Ciafaloni, G.~Curci and A.~Vicere,
  ``Radiative correction effects of a very heavy top,''
  Phys.\ Lett.\ B {\bf 288}, 95 (1992)
  Erratum: [Phys.\ Lett.\ B {\bf 312}, 511 (1993)]
  [hep-ph/9205238].

\bibitem{Fanchiotti:1992tu} 
  S.~Fanchiotti, B.~A.~Kniehl and A.~Sirlin,
  ``Incorporation of QCD effects in basic corrections of the electroweak theory,''
  Phys.\ Rev.\ D {\bf 48}, 307 (1993)
  [hep-ph/9212285].

\bibitem{Djouadi:1993ss} 
  A.~Djouadi and P.~Gambino,
  ``Electroweak gauge bosons selfenergies: Complete QCD corrections,''
  Phys.\ Rev.\ D {\bf 49}, 3499 (1994)
  Erratum: [Phys.\ Rev.\ D {\bf 53}, 4111 (1996)]
  [hep-ph/9309298].

\bibitem{Fleischer:1993ub} 
  J.~Fleischer, O.~V.~Tarasov and F.~Jegerlehner,
  ``Two loop heavy top corrections to the rho parameter: A Simple formula valid for arbitrary Higgs mass,''
  Phys.\ Lett.\ B {\bf 319}, 249 (1993).
  
\bibitem{Avdeev:1994db} 
  L.~Avdeev, J.~Fleischer, S.~Mikhailov and O.~Tarasov,
  ``$O (\alpha \alpha_s^2)$ correction to the electroweak rho parameter,''
  Phys.\ Lett.\ B {\bf 336}, 560 (1994)
  [Phys.\ Lett.\ B {\bf 349}, 597 (1995)]
  [hep-ph/9406363].

\bibitem{Hempfling:1994ar}
  R.~Hempfling and B.~A.~Kniehl,
  ``On the relation between the fermion pole mass and MS Yukawa
  coupling in the standard model,''
  Phys.\ Rev.\ D {\bf 51}, 1386 (1995)
  [hep-ph/9408313].

\bibitem{Larin:1994va} 
  S.~A.~Larin, T.~van Ritbergen and J.~A.~M.~Vermaseren,
  ``The Large quark mass expansion of $\Gamma (Z^0 \rightarrow {\rm hadrons})$ and 
  $\Gamma (\tau^- \rightarrow \nu_\tau + {\rm hadrons})$ in the order $\alpha_s^3$,''
  Nucl.\ Phys.\ B {\bf 438}, 278 (1995)
  [hep-ph/9411260].

\bibitem{Chetyrkin:1995ix} 
  K.~G.~Chetyrkin, J.~H.~Kuhn and M.~Steinhauser,
  ``Corrections of order ${\cal O}(G_F M_t^2 \alpha_s^2)$ to the $\rho$ parameter,''
  Phys.\ Lett.\ B {\bf 351}, 331 (1995)
  [hep-ph/9502291].

\bibitem{Chetyrkin:1995js} 
  K.~G.~Chetyrkin, J.~H.~Kuhn and M.~Steinhauser,
  ``QCD corrections from top quark to relations between electroweak parameters to order alpha-s**2,''
  Phys.\ Rev.\ Lett.\  {\bf 75}, 3394 (1995)
  [hep-ph/9504413].

\bibitem{Degrassi:1996mg} 
  G.~Degrassi, P.~Gambino and A.~Vicini,
  ``Two loop heavy top effects on the m(Z) - m(W) interdependence,''
  Phys.\ Lett.\ B {\bf 383}, 219 (1996)
  [hep-ph/9603374].

\bibitem{Chetyrkin:1997un} 
  K.~G.~Chetyrkin, B.~A.~Kniehl and M.~Steinhauser,
  ``Decoupling relations to $O(\alpha_s^3)$ and their connection to low-energy theorems,''
  Nucl.\ Phys.\ B {\bf 510}, 61 (1998)
  [hep-ph/9708255].

\bibitem{Erler:1998sy} 
  J.~Erler,
  ``Calculation of the QED coupling alpha (M(Z)) in the modified minimal subtraction scheme,''
  Phys.\ Rev.\ D {\bf 59}, 054008 (1999)
  [hep-ph/9803453].

\bibitem{Melnikov:2000qh} 
  K.~Melnikov and T.~v.~Ritbergen,
  ``The Three loop relation between the MS-bar and the pole quark masses,''
  Phys.\ Lett.\ B {\bf 482}, 99 (2000)
  [hep-ph/9912391].

\bibitem{Chetyrkin:2000yt} 
  K.~G.~Chetyrkin, J.~H.~Kuhn and M.~Steinhauser,
  ``RunDec: A Mathematica package for running and decoupling of the strong coupling and quark masses,''
  Comput.\ Phys.\ Commun.\  {\bf 133}, 43 (2000)
  [hep-ph/0004189].

\bibitem{Freitas:2000gg} 
  A.~Freitas, W.~Hollik, W.~Walter and G.~Weiglein,
  ``Complete fermionic two loop results for the M(W) - M(Z) interdependence,''
  Phys.\ Lett.\ B {\bf 495}, 338 (2000)
  Erratum: [Phys.\ Lett.\ B {\bf 570}, no. 3-4, 265 (2003)]
  [hep-ph/0007091].

\bibitem{vanderBij:2000cg} 
  J.~J.~van der Bij, K.~G.~Chetyrkin, M.~Faisst, G.~Jikia and T.~Seidensticker,
  ``Three loop leading top mass contributions to the rho parameter,''
  Phys.\ Lett.\ B {\bf 498}, 156 (2001)
  [hep-ph/0011373].

\bibitem{Jegerlehner:2001fb} 
  F.~Jegerlehner, M.~Y.~Kalmykov and O.~Veretin,
  ``MS versus pole masses of gauge bosons: Electroweak bosonic two loop corrections,''
  Nucl.\ Phys.\ B {\bf 641}, 285 (2002)
  [hep-ph/0105304].

\bibitem{Freitas:2002ja} 
  A.~Freitas, W.~Hollik, W.~Walter and G.~Weiglein,
  ``Electroweak two loop corrections to the $M_W-M_Z$ mass correlation in the standard model,''
  Nucl.\ Phys.\ B {\bf 632}, 189 (2002)
  Erratum: [Nucl.\ Phys.\ B {\bf 666}, 305 (2003)]
  [hep-ph/0202131].

\bibitem{Awramik:2002wn} 
  M.~Awramik and M.~Czakon,
  ``Complete two loop bosonic contributions to the muon lifetime in the standard model,''
  Phys.\ Rev.\ Lett.\  {\bf 89}, 241801 (2002)
  [hep-ph/0208113].

\bibitem{Onishchenko:2002ve} 
  A.~Onishchenko and O.~Veretin,
  ``Two loop bosonic electroweak corrections to the muon lifetime and M(Z) - M(W) interdependence,''
  Phys.\ Lett.\ B {\bf 551}, 111 (2003)
  [hep-ph/0209010].

\bibitem{Jegerlehner:2002em}
F.~Jegerlehner, M.~Y.~Kalmykov and O.~Veretin,
  ``MS-bar versus pole masses of gauge bosons.
  2. Two loop electroweak fermion corrections,''
  Nucl.\ Phys.\ B {\bf 658}, 49 (2003)
  [hep-ph/0212319].

\bibitem{Faisst:2003px}
  M.~Faisst, J.~H.~Kuhn, T.~Seidensticker and O.~Veretin,
  ``Three loop top quark contributions to the rho parameter,''
  Nucl.\ Phys.\ B {\bf 665}, 649 (2003)
 [hep-ph/0302275].

\bibitem{Awramik:2003ee} 
  M.~Awramik and M.~Czakon,
  ``Complete two loop electroweak contributions to the muon lifetime in the standard model,''
  Phys.\ Lett.\ B {\bf 568}, 48 (2003)
  [hep-ph/0305248].

\bibitem{Degrassi:2003rw} 
  G.~Degrassi and A.~Vicini,
  ``Two loop renormalization of the electric charge in the standard model,''
  Phys.\ Rev.\ D {\bf 69}, 073007 (2004)
  [hep-ph/0307122].

\bibitem{Jegerlehner:2003py} 
  F.~Jegerlehner and M.~Y.~Kalmykov,
  ``O(alpha alpha(s)) correction to the pole mass of the t quark within the standard model,''
  Nucl.\ Phys.\ B {\bf 676}, 365 (2004)
  [hep-ph/0308216].

\bibitem{Jegerlehner:2003sp}
  F.~Jegerlehner and M.~Y.~Kalmykov,
  ``O(alpha alpha(s)) relation between pole- and MS-bar mass of the t quark,''
  Acta Phys.\ Polon.\ B {\bf 34}, 5335 (2003)
  [hep-ph/0310361].
  
\bibitem{Awramik:2003rn}
M.~Awramik, M.~Czakon, A.~Freitas and G.~Weiglein,
``Precise prediction for the W boson mass in the standard model,''
Phys. Rev. D \textbf{69}, 053006 (2004)
[arXiv:hep-ph/0311148 [hep-ph]].

\bibitem{Faisst:2004gn}
  M.~Faisst, J.~H.~Kuhn and O.~Veretin,
  ``Pole versus MS mass definitions in the electroweak theory,''
  Phys.\ Lett.\ B {\bf 589}, 35 (2004)
  [hep-ph/0403026].

\bibitem{Kniehl:2004hfa} 
  B.~A.~Kniehl, J.~H.~Piclum and M.~Steinhauser,
  ``Relation between bottom-quark MS-bar Yukawa coupling and pole mass,''
  Nucl.\ Phys.\ B {\bf 695}, 199 (2004)
  [hep-ph/0406254].

\bibitem{Schroder:2005db} 
  Y.~Schroder and M.~Steinhauser,
  ``Four-loop singlet contribution to the rho parameter,''
  Phys.\ Lett.\ B {\bf 622}, 124 (2005)
  [hep-ph/0504055].

\bibitem{Schroder:2005hy} 
  Y.~Schroder and M.~Steinhauser,
  ``Four-loop decoupling relations for the strong coupling,''
  JHEP {\bf 0601}, 051 (2006)
  [hep-ph/0512058].

\bibitem{Chetyrkin:2005ia} 
  K.~G.~Chetyrkin, J.~H.~Kuhn and C.~Sturm,
  ``QCD decoupling at four loops,''
  Nucl.\ Phys.\ B {\bf 744}, 121 (2006)
  [hep-ph/0512060].

\bibitem{Eiras:2005yt}
  D.~Eiras and M.~Steinhauser,
  ``Two-loop O(alpha alpha(s)) corrections to the on-shell
  fermion propagator in the standard model,''
  JHEP {\bf 0602}, 010 (2006)
  [hep-ph/0512099].

\bibitem{Chetyrkin:2006bj} 
  K.~G.~Chetyrkin, M.~Faisst, J.~H.~Kuhn, P.~Maierhofer and C.~Sturm,
  ``Four-Loop QCD Corrections to the Rho Parameter,''
  Phys.\ Rev.\ Lett.\  {\bf 97}, 102003 (2006)
  [hep-ph/0605201].

\bibitem{Boughezal:2006xk} 
  R.~Boughezal and M.~Czakon,
  ``Single scale tadpoles and O(G(F m(t)**2 alpha(s)**3)) corrections to the rho parameter,''
  Nucl.\ Phys.\ B {\bf 755}, 221 (2006)
  [hep-ph/0606232].

\bibitem{Bekavac:2007tk} 
  S.~Bekavac, A.~Grozin, D.~Seidel and M.~Steinhauser,
  ``Light quark mass effects in the on-shell renormalization constants,''
  JHEP {\bf 0710}, 006 (2007)
  [arXiv:0708.1729 [hep-ph]].

\bibitem{Grozin:2011nk} 
  A.~G.~Grozin, M.~Hoeschele, J.~Hoff, M.~Steinhauser, M.~Hoschele, J.~Hoff and M.~Steinhauser,
  ``Simultaneous decoupling of bottom and charm quarks,''
  JHEP {\bf 1109}, 066 (2011)
  [arXiv:1107.5970 [hep-ph]].

\bibitem{Schmidt:2012az} 
  B.~Schmidt and M.~Steinhauser,
  ``CRunDec: a C++ package for running and decoupling of the strong coupling and quark masses,''
  Comput.\ Phys.\ Commun.\  {\bf 183}, 1845 (2012)
  [arXiv:1201.6149 [hep-ph]].

\bibitem{Bezrukov:2012sa}
F.~Bezrukov, M.~Y.~Kalmykov, B.~A.~Kniehl and M.~Shaposhnikov,
``Higgs Boson Mass and New Physics,''
JHEP \textbf{10}, 140 (2012)
[arXiv:1205.2893 [hep-ph]].

\bibitem{Degrassi:2012ry}
G.~Degrassi, S.~Di Vita, J.~Elias-Miro, J.~R.~Espinosa, G.~F.~Giudice, G.~Isidori and A.~Strumia,
``Higgs mass and vacuum stability in the Standard Model at NNLO,''
JHEP \textbf{08}, 098 (2012)
[arXiv:1205.6497 [hep-ph]].

\bibitem{Jegerlehner:2012kn}
  F.~Jegerlehner, M.~Y.~Kalmykov and B.~A.~Kniehl,
  ``On the difference between the pole and the $\overline{\mbox{MS}}$ masses 
  of the top quark at the electroweak scale,''
  Phys.\ Lett.\ B {\bf 722}, 123 (2013)
  [1212.4319].

\bibitem{Buttazzo:2013uya}
D.~Buttazzo, G.~Degrassi, P.~P.~Giardino, G.~F.~Giudice, F.~Sala, A.~Salvio and A.~Strumia,
``Investigating the near-criticality of the Higgs boson,''
JHEP \textbf{12}, 089 (2013)
[arXiv:1307.3536 [hep-ph]].

\bibitem{Kniehl:2014yia} 
  B.~A.~Kniehl and O.~L.~Veretin,
  ``Two-loop electroweak threshold corrections to the bottom and top Yukawa couplings,''
  Nucl.\ Phys.\ B {\bf 885}, 459 (2014)
  Erratum: [Nucl.\ Phys.\ B {\bf 894}, 56 (2015)]
  [arXiv:1401.1844 [hep-ph]].

\bibitem{Bednyakov:2014fua} 
  A.~V.~Bednyakov,
  ``On the electroweak contribution to the matching of the strong coupling constant in the SM,''
  Phys.\ Lett.\ B {\bf 741}, 262 (2015)
  [arXiv:1410.7603 [hep-ph]].

\bibitem{Degrassi:2014sxa} 
  G.~Degrassi, P.~Gambino and P.~P.~Giardino,
  ``The $m_{\scriptscriptstyle W}-m_{\scriptscriptstyle Z}$ interdependence in the Standard Model: a new scrutiny,''
  JHEP {\bf 1505}, 154 (2015)
  [arXiv:1411.7040 [hep-ph]].

\bibitem{Marquard:2015qpa} 
  P.~Marquard, A.~V.~Smirnov, V.~A.~Smirnov and M.~Steinhauser,
  ``Quark Mass Relations to Four-Loop Order in Perturbative QCD,''
  Phys.\ Rev.\ Lett.\  {\bf 114}, no. 14, 142002 (2015)
  [arXiv:1502.01030 [hep-ph]].

\bibitem{Liu:2015fxa} 
  T.~Liu and M.~Steinhauser,
  ``Decoupling of heavy quarks at four loops and effective Higgs-fermion coupling,''
  Phys.\ Lett.\ B {\bf 746}, 330 (2015)
  [arXiv:1502.04719 [hep-ph]].
  
\bibitem{Kniehl:2015nwa} 
  B.~A.~Kniehl, A.~F.~Pikelner and O.~L.~Veretin,
  ``Two-loop electroweak threshold corrections in the Standard Model,''
  Nucl.\ Phys.\ B {\bf 896}, 19 (2015)
  [arXiv:1503.02138 [hep-ph]].

\bibitem{Kniehl:2016enc}
B.~A.~Kniehl, A.~F.~Pikelner and O.~L.~Veretin,
``mr: a C++ library for the matching and running of the Standard Model parameters,''
Comput. Phys. Commun. \textbf{206}, 84-96 (2016)
[arXiv:1601.08143 [hep-ph]].

\bibitem{Marquard:2016dcn} 
  P.~Marquard, A.~V.~Smirnov, V.~A.~Smirnov, M.~Steinhauser and D.~Wellmann,
  ``$\overline{\rm MS}$-on-shell quark mass relation up to four loops in QCD and a general SU$(N)$ gauge group,''
  Phys.\ Rev.\ D {\bf 94}, no. 7, 074025 (2016)
  [arXiv:1606.06754 [hep-ph]].

\bibitem{Bednyakov:2016onn} 
  A.~V.~Bednyakov, B.~A.~Kniehl, A.~F.~Pikelner and O.~L.~Veretin,
  ``On the $b$-quark running mass in QCD and the SM,''
  Nucl.\ Phys.\ B {\bf 916}, 463 (2017)
  [arXiv:1612.00660 [hep-ph]].

\bibitem{Herren:2017osy} 
  F.~Herren and M.~Steinhauser,
  ``Version 3 of RunDec and CRunDec,''
  Comput.\ Phys.\ Commun.\  {\bf 224}, 333 (2018)
  [arXiv:1703.03751 [hep-ph]].
  
\bibitem{Huang:2020hdv}
G.~y.~Huang and S.~Zhou,
``Precise Values of Running Quark and Lepton Masses in the Standard Model,''
Phys. Rev. D \textbf{103}, no.1, 016010 (2021)
[arXiv:2009.04851 [hep-ph]].


\bibitem{Martin:2019lqd}
S.~P.~Martin and D.~G.~Robertson,
``Standard model parameters in the tadpole-free pure $\overline{\rm{MS}}$ scheme,''
Phys. Rev. D \textbf{100}, no.7, 073004 (2019)
[arXiv:1907.02500 [hep-ph]].

\bibitem{SMDRWWW}
The {\tt SMDR} code can be downloaded from:
\href{https://davidgrobertson.github.io/SMDR/}{https://davidgrobertson.github.io/SMDR/}
\\
or \href{https://www.niu.edu/spmartin/SMDR/}{https://www.niu.edu/spmartin/SMDR/}

\bibitem{Martin:2014cxa}
S.~P.~Martin and D.~G.~Robertson,
``Higgs boson mass in the Standard Model at two-loop order and beyond,''
Phys. Rev. D \textbf{90}, no.7, 073010 (2014)
[arXiv:1407.4336 [hep-ph]].

\bibitem{Martin:2015lxa}
S.~P.~Martin,
``Pole Mass of the W Boson at Two-Loop Order in the Pure $\overline {MS}$ Scheme,''
Phys. Rev. D \textbf{91}, no.11, 114003 (2015)
[arXiv:1503.03782 [hep-ph]].

\bibitem{Martin:2015rea}
S.~P.~Martin,
``$Z$-Boson Pole Mass at Two-Loop Order in the Pure $\overline{MS}$ Scheme,''
Phys. Rev. D \textbf{92}, no.1, 014026 (2015)
doi:10.1103/PhysRevD.92.014026
[arXiv:1505.04833 [hep-ph]].

\bibitem{Martin:2016xsp}
S.~P.~Martin,
``Top-quark pole mass in the tadpole-free $\overline {MS}$ scheme,''
Phys. Rev. D \textbf{93}, no.9, 094017 (2016)
[arXiv:1604.01134 [hep-ph]].

\bibitem{Martin:2018yow}
S.~P.~Martin,
``Matching relations for decoupling in the Standard Model at two loops and beyond,''
Phys. Rev. D \textbf{99}, no.3, 033007 (2019)
[arXiv:1812.04100 [hep-ph]].

\bibitem{Martin:2022qiv}
S.~P.~Martin,
``Three-loop QCD corrections to the electroweak boson masses,''
Phys. Rev. D \textbf{106}, no.1, 013007 (2022)
[arXiv:2203.05042 [hep-ph]].

\bibitem{CDF:2022hxs}
T.~Aaltonen \textit{et al.} [CDF],
``High-precision measurement of the W boson mass with the CDF II detector,''
Science \textbf{376}, no.6589, 170-176 (2022)
doi:10.1126/science.abk1781

\end{thebibliography}
\end{document}